%% file: main.tex
\documentclass[
 reprint,
 superscriptaddress,
 nofootinbib,
 amsmath,amssymb,
 aps,prd
]{revtex4-2}

\usepackage{graphicx}
\usepackage{color}
\usepackage{natbib}
\usepackage{epsfig}
\usepackage{float}
\usepackage{xcolor}
\usepackage{soul}
\usepackage{comment}
\usepackage{amsmath,amssymb,amsfonts}
\usepackage{subfigure}
\usepackage{hyperref}
\DeclareUnicodeCharacter{2212}{\textendash}

\input{journalheader}

\begin{document}

\title{Thermal behavior as indicator for hyperons in binary neutron star merger remnants}

\author{Sebastian Blacker}
\affiliation{Institut f\"ur Kernphysik, Technische Universit\"at Darmstadt, 64289 Darmstadt, Germany}
\affiliation{GSI Helmholtzzentrum f\"ur Schwerionenforschung, Planckstra{\ss}e 1, 64291 Darmstadt, Germany}

\author{Hristijan Kochankovski}
\affiliation{Departament de F\'{\i}sica Qu\`antica i Astrof\'{\i}sica and Institut de Ci\`encies del Cosmos, Universitat de Barcelona, Mart\'i i Franqu\`es 1, 08028, Barcelona, Spain}
\affiliation{Faculty of Natural Sciences and Mathematics-Skopje, Ss. Cyril and Methodius University in Skopje, Arhimedova, 1000 Skopje, North Macedonia}

\author{Andreas Bauswein}
\affiliation{GSI Helmholtzzentrum f\"ur Schwerionenforschung, Planckstra{\ss}e 1, 64291 Darmstadt, Germany}
\affiliation{Helmholtz Research Academy Hesse for FAIR (HFHF), Campus Darmstadt, 64291 Darmstadt, Germany}

\author{Angels Ramos}
\affiliation{Departament de F\'{\i}sica Qu\`antica i Astrof\'{\i}sica and Institut de Ci\`encies del Cosmos, Universitat de Barcelona, Mart\'i i Franqu\`es 1, 08028, Barcelona, Spain}

\author{Laura Tolos}
\affiliation{Institute of Space Sciences (ICE, CSIC), Campus UAB, Carrer de Can Magrans, 08193 Barcelona, Spain}
\affiliation{Institut d'Estudis Espacials de Catalunya (IEEC), 08034 Barcelona, Spain}
\affiliation{Frankfurt Institute for Advanced Studies, Ruth-Moufang-Str. 1, 60438 Frankfurt am Main, Germany}
\date{\today}

\begin{abstract}

We provide the first comprehensive study of hyperons in neutron star mergers and quantify their specific impact. We discuss the thermal behavior of hyperonic equations of state~(EoSs) as a distinguishing feature from purely nucleonic models in the remnants of binary mergers using a large set of numerical simulations. Finite temperature enhances the production of hyperons, which leads to a reduced pressure as highly degenerate nucleons are depopulated. This results in a characteristic increase of the dominant postmerger gravitational-wave frequency by up to $\sim150$~Hz compared to purely nucleonic EoS models. By our comparative approach we can directly link this effect to the occurrence of hyperons.
Although this feature is generally weak, it is in principle measurable if the EoS and stellar parameters of cold neutron stars are sufficiently well determined. Considering that the mass-radius relations of purely nucleonic and hyperonic EoSs may be indistinguishable and the overall challenge to infer the presence of hyperons in neutron stars, these findings are important as a new route to answer the outstanding question about hyperonic degrees of freedom in high-density matter.

\end{abstract}


\maketitle   

\section{Introduction}

The ``hyperon puzzle'' states the apparent tension between the expectation that the occurrence of hyperons in dense nuclear matter would soften the equation of state~(EoS) of neutron stars~(NSs) and the observation that some NSs have about two solar masses, which thus requires a certain stiffness of the high-density EoS. Several modern EoS models including hyperons are compatible with these observations (see Refs.~\cite{Chatterjee:2015pua,Oertel:2016bki,Tolos:2020aln,Schaffner-Bielich:2020psc,Burgio:2021vgk,Sedrakian2022,Raduta:2022elz,Sedrakian:2022ata,Takatsuka:2008zz,Bednarek:2011gd,Weissenborn:2011ut,Bonanno:2011ch,Zdunik:2012dj,Yamamoto:2014jga,Lonardoni:2014bwa,Drago:2014oja,Maslov:2015msa,Drago:2015cea,Oertel:2016xsn,Tolos:2016hhl,Tolos:2017lgv,Fortin:2017cvt,Logoteta:2019utx,Ribes:2019kno,Gerstung:2020ktv,Logoteta:2021iuy,Malik:2021nas,Muto:2021jms,Thapa:2021kfo}). This is essentially achieved by tuning some parameters of the phenomenological mean-field models or by including the effect of three-body forces in the microscopical approaches. However, uncertainties remain since both the bare two-body and three-body interactions involving hyperons are poorly known. This is because the available data from scattering experiments are still scarce and subject to quite large error bars, although promising results are becoming available from final-state interaction analyses and femtoscopy studies~\cite{Shapoval:2014yha,HADES:2016dyd,STAR:2014dcy,
CLAS:2021gur,J-PARCE40:2021qxa,J-PARCE40:2021bgw,J-PARCE40:2022nvq,
ALICE:2018ysd,ALICE:2019eol,ALICE:2019buq,
ALICE:2019hdt,ALICE:2021njx,ALICE:2023lgq,ALICE:2020mfd,Fabbietti:2020bfg}.
Hypernuclear structure data can also provide indirect information about the hyperon nuclear forces \cite{Feliciello:2015dua,Tamura:2013lwa,Gal:2016boi}. For instance, recent systematic analyses of $\Lambda$ separation energies in hypernuclei \cite{Friedman:2023ucs}, as well as {\it ab-initio} calculations of light $\Lambda$ hypernuclei employing state-of-the-art chiral interactions \cite{Le:2020zdu,Le:2022ikc,Le:2023bfj}, reveal the need for three-body forces involving hyperons that are expected to have an impact on the EoS around and above saturation density, hence on the properties and composition of neutron stars \cite{Logoteta:2019utx,Gerstung:2020ktv}.
Unfortunately, the many-body treatment also contributes to the total uncertainty, although in the recent decade there has been a significant progress in this area~
\cite{Hebeler:2015hla,Lynn:2019rdt,Hergert:2020bxy,Rios:2020oad,Demol:2019yjt,Drischler:2021kxf,Lovato:2022apd,Arthuis:2022ixv}. The hyperon puzzle thus remains unsolved in the sense that to date it is still not clear whether hyperons are present in NSs.

Unless the problem can be solved self-consistently within many-body theory, i.e.~including the precise knowledge of nucleon and hyperon interactions also at higher densities either from theory or experiment, which seems very ambitious, the solution of the hyperon puzzle necessarily requires the very accurate determination of the mass-radius relation of NSs. From stellar measurements one may either directly infer the presence of hyperons from some specific stellar features in a model-agnostic way or one may gain insights from a detailed comparison between theoretical expectations and observations. As both approaches still appear very challenging, it is justified to consider in this work the case that stellar parameters have been determined with very good precision, which may be achieved in future measurements, and to study which signature can then additionally reveal the presence of hyperons.

\begin{figure}

\includegraphics[width=1.1\linewidth]{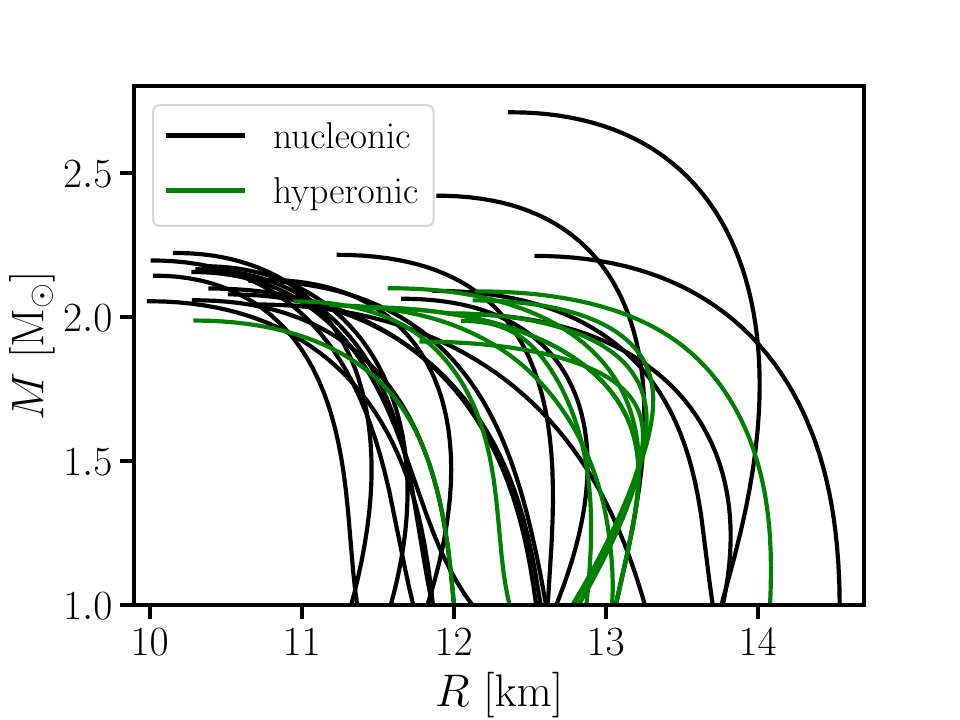}
\caption{Mass-radius relations for purely nucleonic (black) and hyperonic (green) EoSs considered in this study.}
\label{fig:mr}
\end{figure}

Comparing the mass-radius relations of hyperonic and nucleonic EoSs it is difficult to distinguish both classes (see Fig.~\ref{fig:mr}) and hence it is not straightforward to tell from a measured mass-radius relation if the underlying EoS contains hyperons or not. This implies that even when the mass-radius relation of NSs is observationally determined with very good precision, it may still not be possible to deduce the presence of hyperons as no information on the composition can directly be inferred form the stellar parameters. This clearly shows that the identification of hyperonic degrees of freedom in NSs is very difficult and that additional features that may indicate the occurrence of hyperons are highly desirable to solve the hyperon puzzle even for the case that the mass-radius relation is observationally well known.

In this paper we discuss a feature which can be clearly linked to the occurrence of hyperons in NS merger remnants and their associated postmerger gravitational wave (GW) emission: we identify the thermal behavior of hyperonic EoSs as a potential indicator for the presence of hyperons in NS mergers. Using a large set of numerical simulations we present here the first comprehensive study of hyperons in NS mergers that goes beyond comparing individual models (as in~\cite{Sekiguchi:2011mc,Radice:2016rys}), for which it may not be obvious if differences in the observables must be necessarily related to hyperons or could be similarly produced by another nucleonic model with similar EoS properties as the hyperonic model \footnote{As a side note we recall that GN3H and H4 are hyperonic $T=0$ EoS models~\cite{Glendenning:1984jr,Lackey:2005tk}, which have often been used in merger simulations usually as piecewise polytropes~\cite{Read:2008iy} with approximate temperature treatment, e.g.,~\cite{Hotokezaka:2011dh,Bauswein:2012ya,Takami:2014tva,Bernuzzi:2015rla,Dietrich:2016hky,Hanauske:2016gia,Feo:2016cbs,Vretinaris:2019spn,Kedia:2022nns,Raithel:2023zml}. To our knowledge no specific features distinguishing these models (qualitatively) from purely nucleonic models have been reported, which exemplifies the similarity between hyperonic and purely nucleonic models at $T=0$.}.
Ref.~\cite{Sekiguchi:2011mc} reports a difference in the main postmerger GW frequency comparing an EoS model with and without hyperons. In this work we consider all currently available temperature dependent hyperonic EoS models which are publicly accessible and roughly compatible with current astronomical constraints (see e.g.~\cite{Antoniadis:2013pzd,Margalit:2017dij,Bauswein:2017vtn,Shibata:2017xdx,Ruiz:2017due,Radice:2017lry,Most:2018eaw,Rezzolla:2017aly,Koppel:2019pys,LIGOScientific:2018hze,Riley:2019yda,Miller:2019cac,Radice:2018ozg,Coughlin:2018fis,Dietrich:2020efo,Capano:2019eae,Riley:2021pdl,Miller:2021qha,Al-Mamun:2020vzu,Raaijmakers:2021uju,Breschi:2021tbm,Fonseca:2021wxt,Romani:2022jhd,Huth:2021bsp,Huang:2023grj}). We find that the presence of hyperons at finite temperature leads to a small but systematic increase of the dominant postmerger GW frequency up to $\sim150$~Hz compared to purely nucleonic matter.

\section{Methods and setup}

We perform binary merger simulations using a general-relativistic, smoothed particle hydrodynamics~(SPH) code~\cite{Oechslin:2001km,Oechslin:2006uk,Bauswein:2009im} employing the conformal flatness condition to solve the field equations~\cite{Isenberg1980,Wilson:1996ty}. The effects of neutrinos and magnetic fields are not included. We consider equal- and nonequal-mass systems with a total binary mass of $M_\mathrm{tot}=2.8~\mathrm{M}_\odot$. The simulations start with cold, irrotational stars in neutrinoless beta-equilibrium on quasi-equilibrium circular orbits with a center-to-center separation of 38~km. In the simulations we use two samples of EoS models. The first sample contains models that consider the occurrence of different species of hyperons and includes BHB$\Lambda \phi$~\cite{Banik:2014qja}, DD2Y~\cite{Marques:2017zju}, DNS~\cite{Dexheimer:2017nse}, FSU2H*~\cite{Kochankovski:2022rid}, QMC-A~\cite{Stone:2019blq}, R(DD2YDelta)1.1-1.1~\cite{Raduta:2022elz}, R(DD2YDelta)1.2-1.1~\cite{Raduta:2022elz}, R(DD2YDelta)1.2-1.3~\cite{Raduta:2022elz} and SFHOY~\cite{Fortin:2017dsj}. 
We also consider two additional versions of the FSU2H* model, which approximately cover the range of uncertainties of hyperon potentials. Hyperons occur either at lower (FSU2H*L) or higher (FSU2H*U) densities compared to the original FSU2H* model see~\cite{Kochankovski:2023trc} for details). All of these EoSs are developed in the relativistic mean field framework, assuming a strongly repulsive vector meson contribution which is softened by considering density-dependent meson coupling constants or by introducing meson self-interactions. Three of the models, R(DD2YDelta)1.1-1.1, R(DD2YDelta)1.2-1.1 and R(DD2YDelta)1.2-1.3, also consider $\Delta$ resonances as an additional heavy baryon degree of freedom. The impact of $\Delta$ resonances on the EoS is analogous to those of the hyperons. 

The second sample contains purely nucleonic EoSs that do not include exotic degrees of freedom. This sample consists of APR~\cite{Akmal:1998cf,Schneider:2019vdm}, DD2~\cite{Typel:2009sy,Hempel:2009mc}, DD2F~\cite{Typel:2009sy,Alvarez-Castillo:2016oln}, FSU2R~\cite{Tolos:2017lgv}, FTNS~\cite{Furusawa:2017auz,Togashi:2017mjp}, GS2~\cite{Shen:2011kr}, LPB~\cite{Bombaci:2018ksa,Logoteta:2020yxf}, LS220~\cite{Lattimer:1991nc}, LS375~\cite{Lattimer:1991nc}, SFHo~\cite{Hempel:2009mc,Steiner:2012rk}, SFHx~\cite{Hempel:2009mc,Steiner:2012rk}, SRO(SLy4)~\cite{Chabanat:1997un,Schneider:2017tfi}, TM1~\cite{Sugahara:1993wz,Hempel:2011mk} and TMA~\cite{Toki:1995ya,Hempel:2011mk}, and the models 'Fiducial', 'Large Mmax', 'Large SL', 'Large R', 'Small SL' and 'Smaller R' from Ref.~\cite{Du:2021rhq}.

All EoS models are available as fully temperature- and composition-dependent tables. Most of the EoSs are publicly available at the CompOSE website\footnote{https://compose.obspm.fr}~\cite{Typel:2013rza,Oertel:2016bki,CompOSECoreTeam:2022ddl}. Some properties of cold, non-rotating NSs are summarized in Table~\ref{tab:hypresults} for the hyperonic EoS sample and
Table~\ref{tab:nucresults} for the nucleonic one. With the adopted sets, we cover a broad range of stellar parameters roughly compatible with current constraints. All nucleonic EoSs and most of the hyperonic EoSs (except for FSU2H*L) reach a maximum mass above $\approx 2~\mathrm{M}_{\odot}$. 

Note, however, that within the sample of hyperonic EoSs only BHB$\Lambda \Phi$ and DNS are compatible with the recent analysis in~\cite{Romani:2022jhd}, i.e. $M_\mathrm{max}>2.09~\mathrm{M}_\odot$ at the $3\sigma$ confidence level. Also the nucleonic EoSs DD2F, FSU2R, LS220, SFHo, SRO(Sly4) and TMA are in tension with this limit. The nucleonic models GS2, LS375, TM1 and TMA and the hyperonic models DNS, FSU2H*, FSU2H*L and, FSU2H*U are in conflict with the 90\% credible level of the tidal deformability constraint from GW170817~\cite{LIGOScientific:2017vwq,De:2018uhw,LIGOScientific:2018cki}. We include these models in our analysis in order to have a larger sample to be used in simulations.

To analyze the thermal behavior of EoSs, we split the pressure $P$ and specific internal energy $\epsilon$ into a cold and a thermal part, $P=P_\mathrm{cold}+P_\mathrm{th}$ and $\epsilon=\epsilon_\mathrm{cold}+\epsilon_\mathrm{th}$. The thermal components can be related through $P_\mathrm{th}=(\Gamma_\mathrm{th}-1)\epsilon_\mathrm{th}\rho$ adopting a thermal ideal-gas description with the thermal ideal-gas index $\Gamma_\mathrm{th}$ and rest-mass density $\rho$~\cite{Janka1993}. $\Gamma_\mathrm{th}$ is constant for an ideal gas but generally depends on density, temperature and composition for actual microphysical models (see e.g.~\cite{Constantinou:2015mna} for details). The thermal ideal-gas approach can be employed if a temperature extension of microphysical EoS is not available~\cite{Janka1993,Bauswein:2010dn}. The basic approximation consists in choosing a constant $\Gamma_\mathrm{th}$ with a value of $\sim 1.75$ reasonably reproducing the thermal behavior of purely nucleonic microphysical EoSs~\cite{Bauswein:2010dn}.

\section{Approach and results}
Motivated by the fact that mass-radius relations of cold NSs can look very similar for nucleonic and hyperonic EoSs, we focus on the thermal EoS behavior. Therefore, we investigate the dominant postmerger GW frequency $f_\mathrm{peak}$, which in contrast to observables of the binary inspiral phase is affected by finite-temperature effects~\cite{Bauswein:2010dn,Raithel:2021hye,Fields:2023bhs,Raithel:2023zml}. To identify a clear signature of hyperons, we set up a numerical experiment to isolate the impact of the thermal behavior of EoSs.

For this we perform two sets of simulations with all EoS models. First, we run simulations using the full temperature- and composition-dependent EoS tables. For the other set of simulations, we adopt all EoSs at $T=0$ in neutrinoless beta-equilibrium and assume that all these barotropic EoSs would result from purely nucleonic matter\footnote{We here assume that the function $P(\rho)$ for a given cold hyperonic EoS could be similarly produced by different nucleonic interactions within a purely nucleonic model.}. We supplement these barotropic models with the ideal-gas treatment of thermal pressure with $\Gamma_\mathrm{th}=1.75$. This choice mimics the thermal behavior of purely nucleonic matter (as confirmed below), i.e. we equip also the cold hyperonic models with a thermal part characteristic of purely nucleonic matter.

In the simulations using the full EoS table the lepton fraction $Y_{e}$ of each fluid element is determined by cold beta-equilibrium in the setup of the two stars and then advected during the simulation as we do not consider neutrinos. In the simulations employing the barotropic EoS slice with the ideal-gas approach the lepton fraction is always set to the cold beta-equilibrium value at the respective density (see \cite{Bauswein:2010dn} for a discussion). The hyperon content is always assumed to be in weak equilibrium.

To also test asymmetric systems we perform additional simulations with a total binary mass of $2.8~\mathrm{M}_\odot$ and a mass ratio of $q=0.8$ using the hyperonic models DD2Y and FSU2H* and the nucleonic EoSs DD2 and FSU2R with both thermal schemes.

We summarize our simulation results and stellar properties from both sets of simulations in Table~\ref{tab:hypresults} and Table~\ref{tab:nucresults} for the hyperonic and the nucleonic EoS sample, respectively. For hyperonic models we also provide the onset rest-mass density $\rho_\mathrm{onset}$ of hyperons in neutrinoless beta-equilibrium matter at zero temperature as well as the maximum density in the system at the start of the simulation. Underlined values mark systems where hyperons are already present prior to the merger. Tables~\ref{tab:hypresults} and~\ref{tab:nucresults} also report the mass- and time-average values of the thermal ideal-gas index $\bar{\Gamma}_\mathrm{th}$ and the hyperon fraction $\bar{Y}_\mathrm{hyp}$ extracted from the simulations employing the fully temperature-dependent EoSs. For $\bar{\Gamma}_\mathrm{th}$ we first determine a mass-averaged value $\Gamma_\mathrm{th}^\mathrm{av}=\sum m_i \Gamma_{\mathrm{th},i}/\sum m_i$. Here quantities with an index refer to local quantities of a single SPH particle $i$. For particles with a temperature equal to the lowest $T$ in the EoS table (typically 0.1~MeV) we set $\Gamma_{\mathrm{th},i}=1$. We then average $\Gamma_\mathrm{th}^\mathrm{av}$ in a time window of 5~ms starting at 2.5~ms after merger. $\bar{Y}_\mathrm{hyp}$ is calculated analogously.

\begin{table*}
\begin{tabular}{| l | c | c | c | c | c | c | c | c | c | c | c | c |}
\hline
EoS & $M_\mathrm{max}$ & $R_{1.4}$ & $\Lambda_{1.4}$ & $\Lambda_{1.75}$ & $\rho_\mathrm{onset}$  & $\rho_\mathrm{init}^\mathrm{max}$  & $f_\mathrm{peak}$ & $f_\mathrm{peak}$ & $\bar{\Gamma}_\mathrm{th}$ & $\bar{Y}_\mathrm{hyp}$ & $\rho^\mathrm{max}$ & Ref. \\
&  & & & & ($T=0$) &  & (3D) & $(\Gamma_\mathrm{th}=1.75)$ &  & &  & \\
& [$\mathrm{M}_\odot$] & [km] & & & [$10^{15}$~g/cm$^3$] & [$10^{15}$~g/cm$^3$] & [kHz] & [kHz] &  & & [$10^{15}$~g/cm$^3$] & \\ \hline
BHB$\Lambda \phi$ & 2.10 & 13.21 & 695.2 & 160.1 & 0.56 & \underline{0.59} & 2.76 & 2.68 & 1.37 & 0.018 & 0.79 & \cite{Banik:2014qja} \\ \hline
DD2Y & 2.03 & 13.21 & 694.8 & 150.9 & 0.56 & \underline{0.60} & 2.82 & 2.73 & 1.08 & 0.022 & 0.80 & \cite{Marques:2017zju} \\ \hline
DD2Y(q=0.8) & 2.03 & 13.21 & 694.8 & 150.9 & 0.56 & \underline{0.68} & 2.76 & 2.63 & 1.04 & 0.050 & 1.00 & \cite{Marques:2017zju} \\ \hline
DNS & 2.09 & 14.04 & 957.7 & 208.3 & 0.77 & 0.55 & 2.51 & 2.54 & 1.69 & 0.003 & 0.66 & \cite{Dexheimer:2017nse} \\ \hline
FSU2H* & 2.01 & 13.18 & 778.8 & 192.1 & 0.57 & 0.55 & 2.63 & 2.59 & 1.52 & 0.012 & 0.75 & \cite{Kochankovski:2022rid} \\ \hline
FSU2H*(q=0.8) & 2.01 & 13.18 & 778.8 & 192.1 & 0.57 & \underline{0.60} & 2.76 & 2.69 & 1.37 & 0.025 & 0.87 & \cite{Kochankovski:2022rid} \\ \hline
FSU2H*L & 1.91 & 13.16 & 784.4 & 177.6 & 0.56 & 0.54 & 2.68 & 2.62 & 1.24 & 0.018 & 0.76 & \cite{Kochankovski:2022rid,Kochankovski:2023trc} \\ \hline
FSU2H*U & 2.06 & 13.17 & 784.4 & 205.7 & 0.58 & 0.54 & 2.62 & 2.56 & 1.51 & 0.008 & 0.70 & \cite{Kochankovski:2022rid,Kochankovski:2023trc} \\ \hline
QMC-A & 1.99 & 12.89 & 574.8 & 126.0 & 0.93 & 0.66 & 2.91 & 2.98 & 1.65 & 0.003 & 0.91 & \cite{Stone:2019blq} \\ \hline
R(DD2YDelta)1.1-1.1 &  2.04 & 12.96 & 586.8 & 114.0 & 0.46 & \underline{0.69} & 3.03 & 2.93 & 1.08 & 0.083 & 0.95 & \cite{Raduta:2022elz} \\ \hline
R(DD2YDelta)1.2-1.1 & 2.05 & 12.27 & 397.3 & 85.4 & 0.37 & \underline{0.77} & 3.26 & 3.14 & 1.18 & 0.185 & 1.16 & \cite{Raduta:2022elz} \\ \hline
R(DD2YDelta)1.2-1.3 & 2.03 & 13.21 & 696.1 & 150.8 & 0.56 & \underline{0.60} & 2.82 & 2.72 & 0.99 & 0.029 & 0.84 & \cite{Raduta:2022elz} \\ \hline
SFHOY & 1.99 & 11.89 & 333.6 & 61.9 & 0.97 & 0.85 & 3.60 & 3.46 & 1.38 & 0.015 & 1.54 & \cite{Fortin:2017dsj} \\ \hline
\end{tabular}
\caption{Sample of EoSs considered in this work which include hyperonic degrees of freedom. Second to fifth column provide properties of cold stars, namely the maximum mass $M_\mathrm{max}$, the radius $R_{1.4}$ and tidal deformability $\Lambda_{1.4}$ of a $1.4~\mathrm{M}_\odot$ neutron star and the tidal deformability $\Lambda_{1.75}$ of a $1.75~\mathrm{M}_\odot$ neutron star. $\rho_\mathrm{onset}$ is the onset rest-mass density for the occurrence of hyperons in the $T=0$, beta-equilibrium EoS slice and $\rho_\mathrm{init}^\mathrm{max}$ corresponds to the maximum density in the system at the beginning of the simulation. Underlined values mark systems with hyperons present prior to the merger. Next two columns report the dominant postmerger GW frequency $f_\mathrm{peak}$ from simulations using either the full temperature-dependent EoS table or the barotropic EoS table together with the ideal-gas approximation for thermal pressure with $\Gamma_\mathrm{th}=1.75$. $\bar{\Gamma}_\mathrm{th}$ and $\bar{Y}_\mathrm{hyp}$ refer to the mass- and time-averaged thermal ideal-gas index and hyperon fraction of the remnant, respectively. $\rho^\mathrm{max}$ is the maximum rest-mass density within the first 5~ms after merger.}
\label{tab:hypresults}
\end{table*}

\begin{table*}
\begin{tabular}{| l | c | c | c | c | c | c | c | c | c |}
\hline\
EoS & $M_\mathrm{max}$ & $R_{1.4}$ & $\Lambda_{1.4}$ & $\Lambda_{1.75}$ & $f_\mathrm{peak}\mathrm{(3D)}$ & $f_\mathrm{peak}(\Gamma_\mathrm{th}=1.75)$ & $\bar{\Gamma}_\mathrm{th}$ & $\rho^\mathrm{max}$ & Ref. \\
& [$\mathrm{M}_\odot$] & [km] & & & [kHz] & [kHz] & & [$10^{15}$~g/cm$^3$] & \\ \hline
APR & 2.20 & 11.57 & 267.6 & 54.5 & 3.51 & 3.46 & 1.74 & 1.41 & \cite{Akmal:1998cf,Schneider:2019vdm}\\ \hline
DD2 & 2.42 & 13.22 & 698.8 & 178.5 & 2.64 & 2.68 & 1.78 & 0.71 & \cite{Typel:2009sy,Hempel:2009mc}\\ \hline
DD2(q=0.8) & 2.42 & 13.22 & 698.8 & 178.5 & 2.68 & 2.69 & 1.74 & 0.73 & \cite{Typel:2009sy,Hempel:2009mc}\\ \hline
DD2F & 2.08 & 12.40 & 425.5 & 79.3 & 3.30 & 3.30 & 1.66 & 1.12 & \cite{Typel:2009sy,Alvarez-Castillo:2016oln}\\ \hline
DSH Fiducial & 2.17 & 11.73 & 296.3 & 61.8 & 3.44 & 3.40 & 1.77 & 1.28 & \cite{Du:2021rhq}\\ \hline
DSH Large Mmax & 2.22 & 12.65 & 513.9 & 119.9 & 2.93 & 2.91 & 1.79 & 0.85 & \cite{Du:2021rhq}\\ \hline
DSH Large SL & 2.16 & 11.76 & 271.5 & 55.9 & 3.51 & 3.46 & 1.52 & 1.38 & \cite{Du:2021rhq}\\ \hline
DSH Large R & 2.13 & 12.44 & 437.6 & 87.3 & 3.16 & 3.18 & 1.72 & 1.08 & \cite{Du:2021rhq}\\ \hline
DSH Small SL & 2.18 & 11.70 & 335.8 & 70.3 & 3.31 & 3.33 & 1.76 & 1.21 & \cite{Du:2021rhq}\\ \hline
DSH Smaller R & 2.14 & 11.29 & 233.1 & 48.8 & 3.62 & 3.60 & 1.72 & 1.66 & \cite{Du:2021rhq}\\ \hline
FSU2R & 2.06 & 12.87 & 640.8 & 143.5 & 2.80 & 2.81 & 1.81 & 0.83 & \cite{Tolos:2017lgv}\\ \hline
FSU2R(q=0.8) & 2.06 & 12.87 & 640.8 & 143.5 & 2.69 & 2.70 & 1.76 & 0.91 & \cite{Tolos:2017lgv}\\ \hline
FTNS & 2.22 & 11.46 & 304.8 & 65.3 & 3.34 & 3.40 & 1.73 & 1.26 & \cite{Furusawa:2017auz,Togashi:2017mjp}\\ \hline
GS2 & 2.09 & 13.60 & 721.3 & 160.6 & 2.73 & 2.70 & 1.76 & 0.73 & \cite{Shen:2011kr}\\ \hline
LPB & 2.10 & 12.37 & 429.9 & 79.9 & 3.23 & 3.23 & 1.68 & 1.01 & \cite{Bombaci:2018ksa,Logoteta:2020yxf}\\ \hline
LS220 & 2.04 & 12.96 & 541.9 & 94.2 & 3.09 & 3.06 & 1.54 & 1.00 & \cite{Lattimer:1991nc}\\ \hline
LS375 & 2.71 & 13.95 & 960.1 & 257.7 & 2.44 & 2.44 & 1.63 & 0.59 & \cite{Lattimer:1991nc}\\ \hline
SFHo & 2.06 & 11.89 & 333.5 & 63.5 & 3.43 & 3.45 & 1.62 & 1.42 & \cite{Hempel:2009mc,Steiner:2012rk}\\ \hline
SFHx & 2.13 & 11.98 & 395.1 & 86.7 & 3.16 & 3.18 & 1.82 & 1.09 & \cite{Hempel:2009mc,Steiner:2012rk}\\ \hline
SRO(SLy4) & 2.05 & 11.72 & 303.7 & 54.7 & 3.51 & 3.50 & 1.78 & 1.43 & \cite{Chabanat:1997un,Schneider:2017tfi}\\ \hline
TM1 & 2.21 & 14.47 & 1149.0 & 257.7 & 2.38 & 2.40 & 1.82 & 0.55 & \cite{Sugahara:1993wz,Hempel:2011mk}\\ \hline 
TMA & 2.01 & 13.79 & 929.1 & 184.1 & 2.58 & 2.57 & 1.74 & 0.66 & \cite{Toki:1995ya,Hempel:2011mk}\\ \hline

\hline
\end{tabular}
\caption{Sample of purely nucleonic EoSs considered in this work. Second to fifth column provide properties of cold stars, namely the maximum mass $M_\mathrm{max}$, the radius $R_{1.4}$ and tidal deformability $\Lambda_{1.4}$ of a $1.4~\mathrm{M}_\odot$ neutron star and the tidal deformability $\Lambda_{1.75}$ of a $1.75~\mathrm{M}_\odot$ neutron star. Next two columns report the dominant postmerger GW frequency $f_\mathrm{peak}$ from simulations using either the full temperature-dependent EoS table or the barotropic EoS table together with the ideal-gas approximation for thermal pressure with $\Gamma_\mathrm{th}=1.75$. $\bar{\Gamma}_\mathrm{th}$ refers to the mass- and time-averaged thermal ideal-gas index of the remnant. $\rho^\mathrm{max}$ is the maximum rest-mass density within the first 5~ms after merger.}
\label{tab:nucresults}
\end{table*}

\begin{figure}
\includegraphics[width=1.1\linewidth]{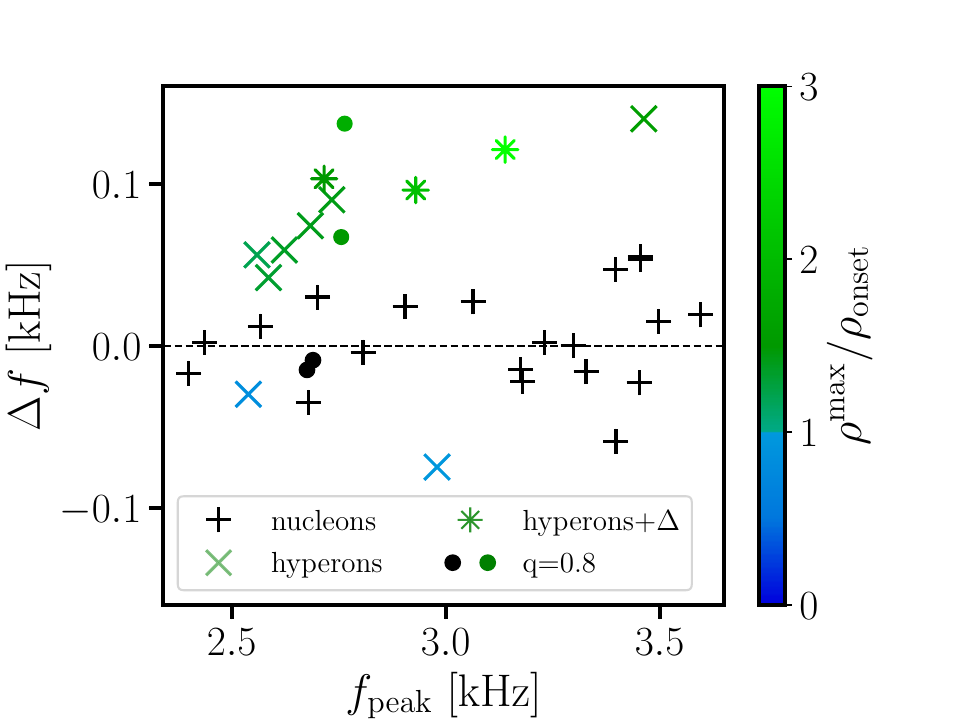}
\caption{Difference $\Delta f=f_\mathrm{peak}-f_\mathrm{peak}^{1.75}$ between dominant postmerger GW frequency of simulations with fully temperature-dependent EoSs and calculations with same EoS models restricted to zero temperature and supplemented with an ideal-gas treatment of thermal effects, which mimics the behavior of purely nucleonic EoSs by choosing a thermal ideal-gas index of $\Gamma_\mathrm{th}=1.75$. Shown as function of $f_\mathrm{peak}$. Black symbols depict purely nucleonic models. Crosses display hyperonic models, where the coloring indicates the ratio between the maximum rest-mass density in the postmerger remnant and the onset rest-mass density of hyperon production at zero temperature. Asterisks refer to models which additionally include $\Delta$-baryons. Circles display results from asymmetric binaries.} 
\label{fig:df}
\end{figure}

We now compare the dominant postmerger GW frequencies of the full models, $f_\mathrm{peak}$, with the frequencies $f_\mathrm{peak}^{1.75}$ obtained from the $\Gamma_\mathrm{th}=1.75$ calculations. The difference $\Delta f \equiv f_\mathrm{peak}-f_\mathrm{peak}^{1.75}$ describes how well the thermal behavior of the full EoS table is modeled by $\Gamma_\mathrm{th}=1.75$ and thus measures by how much a given model deviates from an idealized ``nucleonic'' thermal behavior. The results are given in Fig.~\ref{fig:df}, where we distinguish purely nucleonic models (black) and hyperonic models (colored). As anticipated, the purely nucleonic models cluster around zero corroborating that $\Gamma_\mathrm{th}=1.75$ is a good choice for nucleonic matter (but see discussion below). Hyperonic models lead to systematically higher frequencies compared to the $\Gamma_\mathrm{th}=1.75$ runs, which mimic a nucleonic behavior.

The color of the symbols in Fig.~\ref{fig:df} indicates by how much the maximum rest-mass density in the merger remnant within the first 5~ms after the merger, $\rho^\mathrm{max}$, exceeds the rest-mass density $\rho_\mathrm{onset}$ where hyperons occur in cold matter for the given EoS. Models with $\rho^\mathrm{max}/\rho_\mathrm{onset}<1$ still contain a tiny amount of hyperons due to their occurrence at finite temperature (about 0.3\%; see Tab.~\ref{tab:hypresults}). 
The coloring in Fig.~\ref{fig:df} looks very similar if we choose it to directly indicate the hyperon content in the remnant or a mass-averaged thermal ideal-gas index since these two quantities correlate with $\rho^\mathrm{max}/\rho_\mathrm{onset}$. Obviously, if only a very small amount of hyperons is present, hardly any impact on the dominant postmerger frequency is expected, which is why two blue symbols in Fig.~\ref{fig:df} are located around $\Delta f\sim 0$ as the nucleonic models. We also remark that the results from the asymmetric binaries are in good agreement with the findings from symmetric systems. 

A frequency shift of $f_\mathrm{peak}$ of about 100~Hz is small (compared to the FWHM of the peaks or the variation of $f_\mathrm{peak}$ with the EoS) but potentially sizable enough for a detection. Clearly, a frequency shift itself cannot be measured directly since only one true neutron star EoS exists. We envision a scenario where an observed $f_\mathrm{peak}$ has to be compared to simulation results with different thermal approaches.

In other words, suppose the cold EoS was known with good precision (e.g.~from GW inspiral measurements), but the actual content of the matter - whether hyperons are present or not - remains unknown. This degeneracy is broken by thermal effects, which influence $f_\mathrm{peak}$. One can then perform merger simulations to predict a reference value, $f_\mathrm{peak}^{1.75}$ using the cold EoS and the ideal-gas approach with $\Gamma_\mathrm{th}=1.75$, i.e. a typical thermal behavior for purely nucleonic matter. A deviation between $f_\mathrm{peak}^{1.75}$ 
and the actually measured $f_\mathrm{peak}$ by around 50~Hz to 150~Hz would indicate the presence of hyperons.

Gravitational-wave injection studies as in~\cite{Clark:2014wua,Chatziioannou:2017ixj,Yang:2017xlf,Torres-Rivas:2018svp,Tsang:2019esi,Easter:2020ifj,Breschi:2022ens,Wijngaarden:2022sah,Criswell:2022ewn} show that $f_\mathrm{peak}$ may be determined to within $\sim10$~Hz by future facilities such that a frequency shift of this order is in principle measurable and thus the effect of hyperons would be accessible. 

However, this requires not only the cold EoS to be measured sufficiently well but, additionally, that simulation tools are reliable enough to predict an accurate reference value $f_\mathrm{peak}^{1.75}$ for the comparison with the observational data.
Both prerequisites are currently not given but might be achieved in the future albeit they clearly represent challenging efforts. To provide a coarse estimate of the requirements we note that a frequency shift of 100~Hz corresponds to a change of the NS radius of about 250~m considering empirical relations that connect $f_\mathrm{peak}$ and the radius of cold, nonrotating NSs~\cite{Bauswein:2015vxa}.

Another potentially very promising route to identify the presence of hyperons links two directly measurable quantities.  Importantly, this detection scenario does not assume that the cold EoS is known. Employing the same simulation data, we relate $f_\mathrm{peak}$ and the tidal deformability $\Lambda$. The tidal deformability describing finite-size effects during the GW inspiral is given by $\Lambda=\frac{2}{3}k_2(\frac{R}{M})^5$ with the stellar mass and radius, $M$ and $R$, and the tidal Love number $k_2$ being a function of mass and the EoS~\cite{Flanagan:2007ix,Hinderer:2007mb,Hinderer:2009ca,Damour:2009wj,Damour:2012yf}. As the stellar radius, $\Lambda$ characterizes the cold EoS and it can be measured during the GW inspiral phase (see, e.g.~\cite{LIGOScientific:2017vwq,LIGOScientific:2018hze,Chatziioannou:2020pqz,Dietrich:2020eud}).

Given that we consider 1.4~$\mathrm{M}_\odot$-1.4~$\mathrm{M}_\odot$ binaries a natural choice would be to relate $f_\mathrm{peak}$ to the tidal deformability $\Lambda_{1.4}$ of a 1.4~$\mathrm{M}_\odot$ star. However, when relating $f_\mathrm{peak}$ and $\Lambda$, the reference mass $M_\mathrm{ref}$ at which $\Lambda$ is evaluated can be in principle chosen freely but the choice affects the accuracy of the relation~\cite{Lioutas:2021jbl}.

\begin{figure}
\centering
\includegraphics[width=1.1\linewidth]{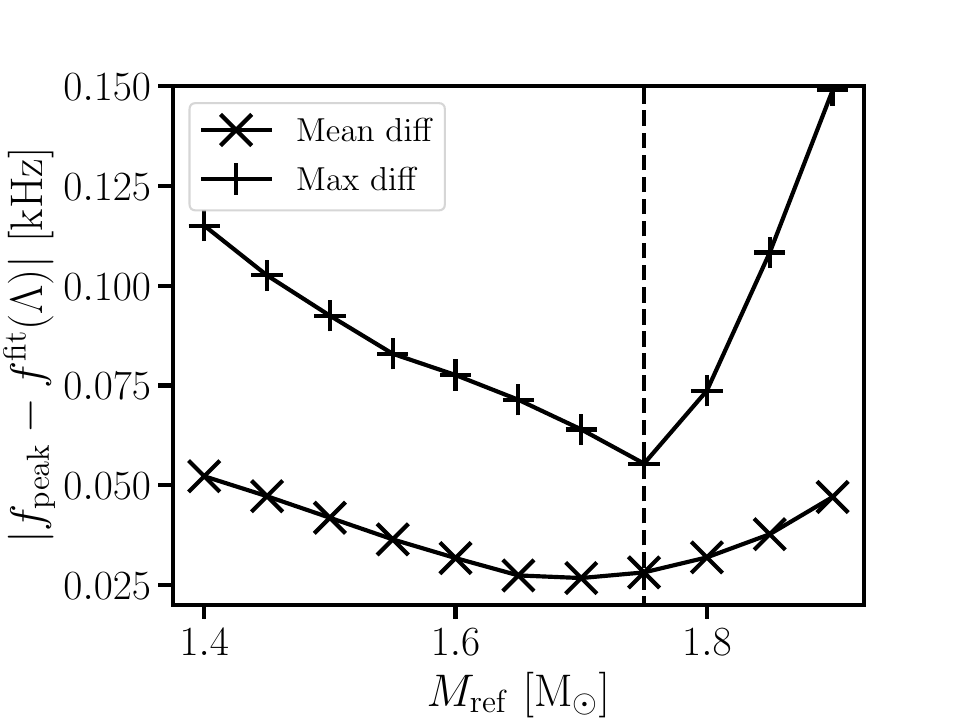}
\caption{Mean and maximum deviation of our data for purely nucleonic EoSs from quadratic $f_\mathrm{peak}-\Lambda_M$ least-squares fits for different reference masses $M_\mathrm{ref}$ at which $\Lambda$ is evaluated ($\Lambda_M=\Lambda(M_\mathrm{ref})$). The dashed vertical line indicates the minimum of the largest deviation at $M_\mathrm{ref}=1.75$.}
\label{fig:flaccuracy}
\end{figure}

Similar to Ref.~\cite{Lioutas:2021jbl}, in Fig.~\ref{fig:flaccuracy} we plot the mean and the maximum deviation of our data for purely nucleonic EoSs from the quadratic $f_\mathrm{peak}-\Lambda_M$ fit for different $M_\mathrm{ref}$ with $\Lambda_M=\Lambda(M_\mathrm{ref})$. As expected we find that the scatter depends on $M_\mathrm{ref}$. The maximum deviation reaches a minimum of 55~Hz at $M_\mathrm{ref}=1.75~\mathrm{M}_\odot$, indicated by the vertical dashed line. Since this relation has the smallest overall scatter, we compare the results from the hyperonic EoS sample to this relation.

\begin{figure}
\includegraphics[width=1.1\linewidth]{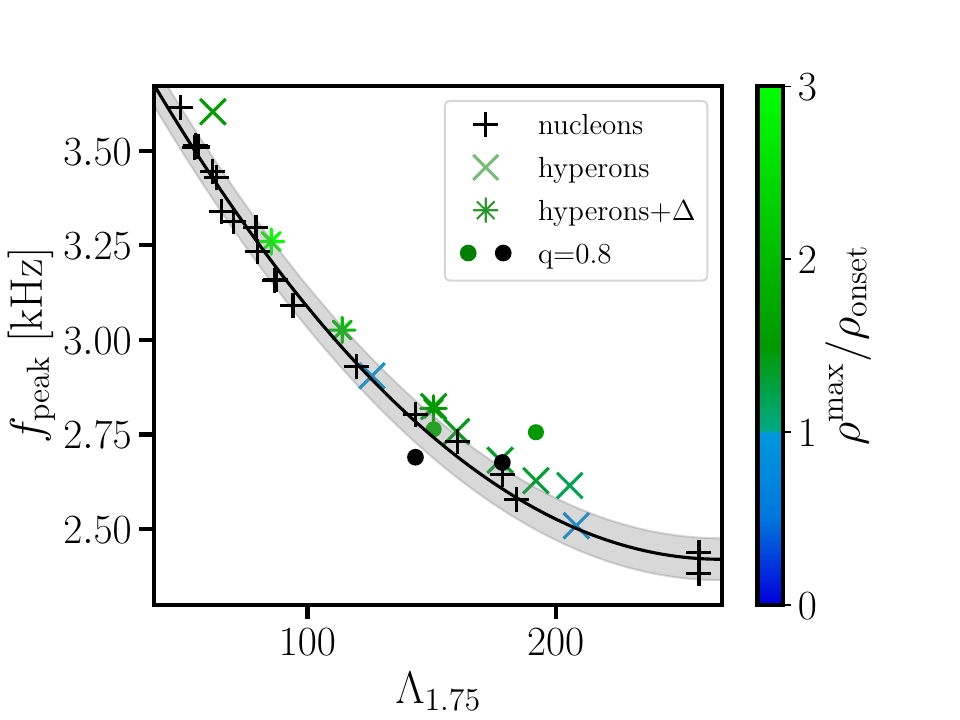}
\caption{Dominant postmerger GW frequency of 1.4~$\mathrm{M}_\odot$-1.4~$\mathrm{M}_\odot$ mergers as function of tidal deformability of a 1.75~$\mathrm{M}_\odot$ NS. Symbols and color scheme as in Fig.~\ref{fig:df}. Black curve shows least-squares fit to purely nucleonic models. Gray band indicates maximum residual of purely nucleonic models from the fit.}
\label{fig:fl}
\end{figure}

In Fig.~\ref{fig:fl} we display $f_\mathrm{peak}$ from the simulations with the fully temperature-dependent EoSs as function of the tidal deformability of 1.75~$\mathrm{M}_\odot$ NSs. The black line is a least-squares quadratic fit to the purely nucleonic models and those data points deviate at most by 55~Hz from the fit. The average deviation of nucleonic models is only 28~Hz. As suggested by Fig.~\ref{fig:df} the dominant postmerger frequency of hyperonic models (colored symbols) is characteristically increased compared to the nucleonic models. Most hyperonic models lie above the frequency range which is spanned by the fit to nucleonic models and the maximum residual of nucleonic models (visualized by the gray band).
This implies that at least in principle the presence of hyperons may be deduced by an increased postmerger frequency which is incompatible with a purely nucleonic EoS. On the other hand, hyperon content up to a certain density~(see Fig.~8 in~\cite{Blacker:2020nlq}) may be excluded if inferred values of $ \Lambda_{1.75}$ and $f_\mathrm{peak}$ lie below the fit to purely nucleonic models. Obviously, this scenario relies on simulations being sufficiently accurate in predicting $f_\mathrm{peak}$ for a given EoS. Again simulations where only a small amount of hyperons is present (blue symbols), lie practically on top of the fit to nucleonic EoSs. For the two asymmetric binaries with nucleonic EoSs we observe a somewhat larger scatter from the relation. This could imply that a different reference mass should be used for these systems or that $f_\mathrm{peak}-\Lambda$ relations are simply not as tight in the case of asymmetric binaries. This should be further investigated in future work.

Some models with a sizable fraction of hyperons do not stick out very clearly due to the overall properties of their EoS. These hyperonic models do result in a significant frequency shift by the thermal behavior of the hyperons, as can be seen in Fig.~\ref{fig:df}. However, this shift essentially only compensates the fact that the fiducial ``nucleonic'' models (i.e. with $\Gamma_\mathrm{th}=1.75$) produce relatively low frequencies in Fig.~\ref{fig:fl}, i.e. close to the lower edge of the gray band. This can be seen by the frequency shift of the respective model in Fig.~\ref{fig:df}. 

This shows that it is not straightforward to connect the exact location of a hyperonic model with respect to the fit in Fig.~\ref{fig:fl} with the actual amount of hyperons in the remnant, albeit one may generally conclude that the presence of hyperons seems more likely if the postmerger frequency is high compared to the fit. The exact location is a superposition of the thermal behavior and properties of the cold EoS which are not captured by $\Lambda_{1.75}$. Our set of EoSs does not represent a statistical ensemble and one should thus be careful with likelihood arguments.

Since the magnitude of the frequency shift of hyperonic models is generally small, the intrinsic scatter of $f_\mathrm{peak}(\Lambda)$ relations is particularly relevant and a better understanding of the scatter of such empirical relations will improve the prospects to infer the presence of hyperons.

\begin{figure}
\includegraphics[width=1.1\linewidth]{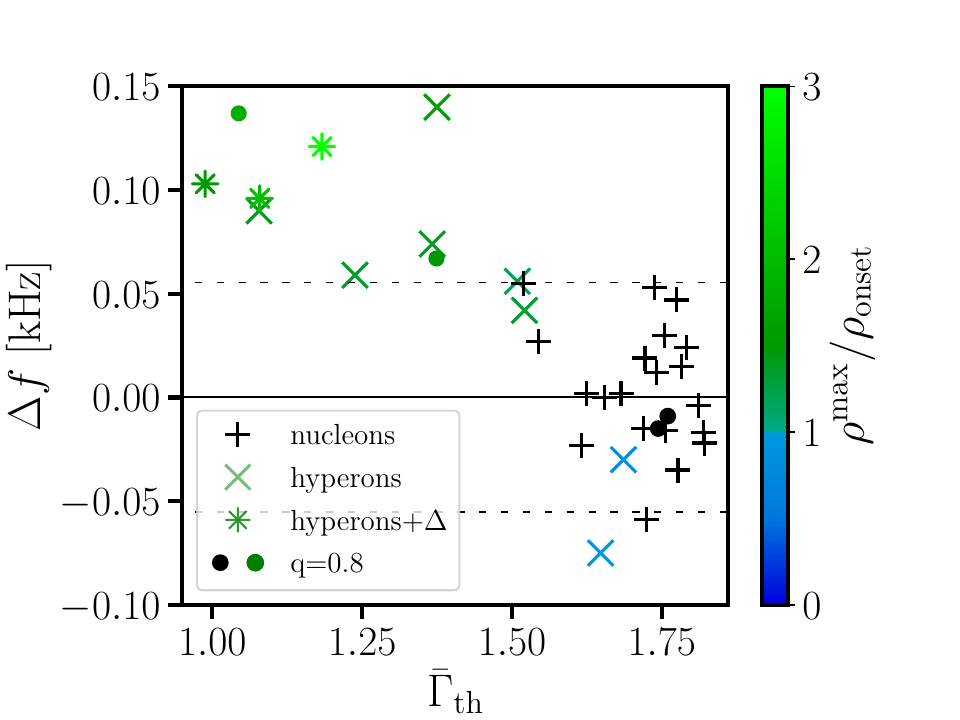}
\caption{Difference $\Delta f=f_\mathrm{peak}-f_\mathrm{peak}^{1.75}$ as function of mass- and time-averaged thermal index of the merger remnant in simulations employing the temperature-dependent EoSs. Same symbols and color scheme as in Fig.~\ref{fig:df}. Dashed lines indicate maximum residual of $f_\mathrm{peak}$-$\Lambda_{1.75}$ fit to purely nucleonic EoSs from Fig.~\ref{fig:fl}.}
\label{fig:dfgth}
\end{figure}

Figure~\ref{fig:dfgth} illustrates the dependence of GW frequency shifts $\Delta f$ on mass-averaged and time-averaged ideal-gas index $\bar{\Gamma}_\mathrm{th}$. As in Fig.~\ref{fig:df} we also include some results for asymmetric binaries. The figure clearly shows that the presence of hyperons leads to a reduction of the thermal pressure component in comparison to purely nucleonic models. The increase of the thermal energy favors the occurrence of hyperons, which, in turn, reduces pressure from the highly degenerate species resulting in a substantial decrease of the thermal index. The figure also summarizes the observation that many of the hyperonic models feature a frequency shift that is larger than the maximum residual of the fit to nucleonic models in Fig.~\ref{fig:fl}. Figure~\ref{fig:dfgth} also demonstrates that $\Gamma_\mathrm{th}=1.75$ is a good choice to model the thermal behavior of nucleonic EoSs. 

\begin{figure}
\centering
\includegraphics[width=1.1\linewidth]{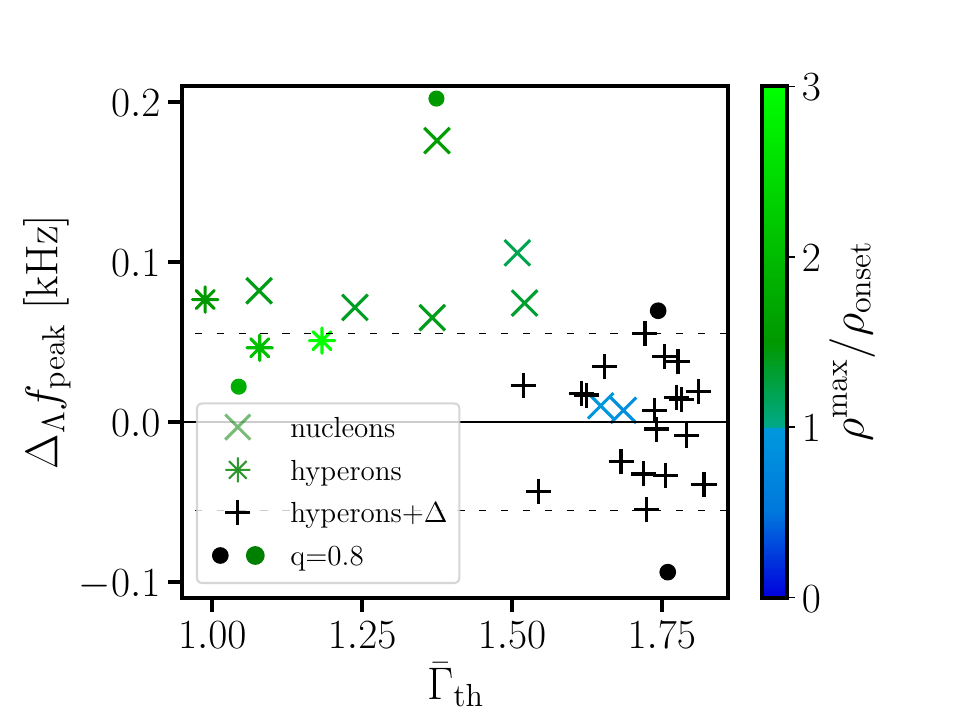}
\caption{Difference between the dominant postmerger GW frequency of 1.4~$\mathrm{M}_\odot$-1.4~$\mathrm{M}_\odot$ mergers and the frequency given by the least-squares fit of $f_\mathrm{peak}(\Lambda_{175})$  for hyperonic and purely nucleonic models (see Fig.~\ref{fig:fl}) as function of the mass- and time averaged thermal ideal-gas index of the remnant. Same symbols and color scheme as in Fig.~\ref{fig:df}. Dashed lines indicate maximum residual of the fit to purely nucleonic EoSs.}
\label{fig:flgth}
\end{figure}

Additionally, in Fig.~\ref{fig:flgth} we visualize the deviations from the $f_\mathrm{peak}-\Lambda_{1.75}$ relation for purely nucleonic models as a function of the mass- and time-averaged thermal ideal-gas index $\Bar{\Gamma}_\mathrm{th}$ for purely nucleonic and hyperonic EoSs. Again, we observe that for most hyperonic EoS models the deviation from the relation is comparable or slightly above the maximum residual of the relation for purely nucleonic models. Since the magnitude of the frequency deviation from the fit is influenced by both the properties of the cold and the finite-temperature EoS, the deviations do not exactly scale with $\Bar{\Gamma}_\mathrm{th}$ or $\rho^\mathrm{max}/\rho_\mathrm{onset}$.

\section{Toy model: Thermal behavior of hyperons}

In recent studies based on chiral effective field theory ($\chi$EFT) for nucleons~\cite{Keller:2020qhx,Keller:2022crb} a drop of the thermal index with density is observed. In these state-of-the art calculations for nuclear matter, the thermal pressure is reduced with density due to the increase of the effective nucleon mass.
This stems from the strong three body force within $\chi$EFT, which in turn reduces the thermal index (see Eq.~(41) in Ref.~\cite{Keller:2020qhx}). However, $\chi$EFT in dense nuclear matter is applicable for densities up to around 2$n_0$ ($n_0$ being the nuclear saturation density) and temperature $T \lesssim 30$ MeV~\cite{Drischler:2021kxf} and, therefore, the resulting EoS cannot be used directly in NS merger simulations. Moreover, up to now, the above $\chi$EFT finite-temperature framework does not include exotic degrees of freedom, such as hyperons.

If the drop of the thermal index as predicted within $\chi$EFT~\cite{Keller:2020qhx,Keller:2022crb} is a generic feature within a larger density range, this property could mimic the thermal behavior of hyperons as discussed in this work. We therefore explore, within a toy model that shows a similar drop of the thermal index in nucleonic matter as that of the phenomenological extrapolation of the microscopic $\chi$EFT model performed in~\cite{Huth:2020ozf}, what would be the additional effect of the hyperonic degrees of freedom. To this end we build a toy model within the relativistic mean-field framework, adopting the DDME2 model~\cite{Lalazissis:2005de} extended to the hyperonic sector with the density-dependent couplings as defined in~\cite{Sedrakian:2022kgj}. We impose a similar drop of the thermal index in purely nucleonic matter as that in~\cite{Huth:2020ozf} by modifying the functional density dependence of the $\sigma$-meson coupling for nucleons, such that their effective mass experiences a minimum close to $n_0$. Other meson couplings remain unchanged. We stress that the EoS constructed in this way is not compatible with nuclear physics constraints, coming from the properties of
nuclear matter, nuclei, heavy ion collisions at high energies and astrophysical observations (see, for example, reviews on the EoS and the nuclear and astrophysical constraints of Refs.~\cite{FiorellaBurgio:2018dga,Tolos:2020aln,Burgio:2021vgk}). This setup just provides a purely nucleonic model with a thermal index sensibly below $\Gamma_\mathrm{th}=4/3$, allowing us to additionally include hyperons and analyze their thermal effects. This is meant to resemble the aforementioned behavior found in the $\chi$EFT framework but with the inclusion of hyperons.

\begin{figure}[h]
    \centering 
    \includegraphics[width=1.1\linewidth]{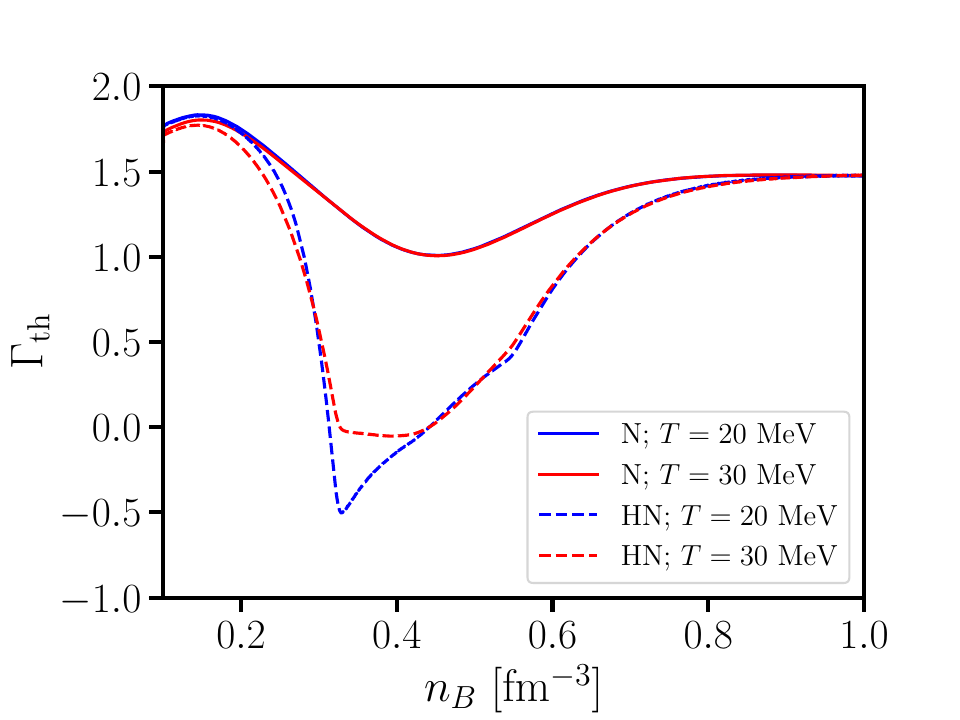}
   \caption{The thermal index for nucleonic (solid lines) and hyperonic matter (dashed lines) for two different temperatures, $T=20$ MeV (blue lines) and $T=30$ MeV (red lines). See text for details of the model. }
    \label{fig:thermal}
\end{figure}

From Fig.~\ref{fig:thermal} one can see that the occurrence of hyperons in $\beta^{-}$ equilibrated matter at densities above $0.2 \ {\rm fm^{-3}}$ gives rise to a significantly stronger reduction of the thermal index than in the nucleonic case. Similarly to the EoS models employed in this work, this additional drop of the thermal index is due to the loss of degeneracy pressure as hyperons appear, a mechanism that is apparently present regardless of the underlying nucleonic interaction. 

However, a few words of caution are in order. As already mentioned, this toy model does not reproduce neither nuclear matter properties nor astrophysical observations, and therefore a more realistic simulation should be investigated in future work. We also note that in the relativistic-mean-field scheme employed to build our toy model the effective masses of hyperons are correlated with the nucleonic ones, as the coupling constants for nucleons and hyperons are related by symmetry relations. This implies that the effective masses of the nucleons and hyperons have the same functional dependence, which is not necessarily the case, and this may have an influence on the size of the additional hyperonic drop. As these questions are beyond the scope of the present work, based on the results of this study we can simply conclude that even if a nucleonic model would have rather low values of the thermal index, one should expect an additional decrease due to the occurrence of hyperons. 

\section{Summary and discussion}
We conclude this work by discussing the prospects of identifying hyperons in NSs through their thermal properties as well as the possible caveats. By the approach chosen in this study we for the first time isolate the specific impact of the thermal behavior of hyperons in NS mergers. Above we already discussed caveats and challenges of identifying hyperons through their thermal behavior like requiring very precise stellar measurements (e.g. of $\Lambda_{1.75}$) and accurate simulations for reference. We have motivated and justified these rather optimistic assumptions by the difficulties to otherwise obtain information about the presence of hyperons in NSs. Any attempt to infer the presence of hyperons through astronomical measurements similarly requires very high or even higher precision to decipher their very weak impact on the stellar structure. Therefore, any new, additional feature which can be linked to hyperons as the one discussed in this paper is highly valuable. A few more comments on the prospects and caveats are in order.

We here explicitly assumed that the properties of the cold EoS and of cold, isolated NSs do not carry any information on the presence of hyperons. This was necessary to neatly quantify the impact of hyperons on the thermal behavior, but this is a very conservative and in fact incorrect assumption. In reality the cold EoS and NS parameters are affected by hyperons and for instance theoretical calculations of purely nucleonic matter may in future already indicate the presence of additional degrees of freedom when compared to measurements of cold NSs. Also advances on the experimental determination of two-body and three-body interactions involving hyperons and nucleons, e.g. at J-PARC, LHC or the future FAIR facility, and corresponding theoretical progress can be anticipated~\cite{CLAS:2021gur,J-PARCE40:2021qxa,J-PARCE40:2021bgw,J-PARCE40:2022nvq,
ALICE:2018ysd,ALICE:2019eol,ALICE:2019buq,
ALICE:2019hdt,ALICE:2021njx,ALICE:2023lgq,Durante:2019hzd,Hebeler:2020ocj,Petschauer:2020urh,Haidenbauer:2023qhf} , which would lead to further constraints that can be incorporated in future analyses. Ab-initio calculations that consider hyperons as relevant degrees of freedom may furthermore constrain the parameter space~\cite{Wirth:2017bpw,Wirth:2019cpp,Le:2020zdu,Le:2022ikc}. Further insights may result from astronomical observations like cooling NSs or core-collapse supernovae~\cite{Prakash:1992zng,Page:2004fy,Yakovlev:2004iq,Sumiyoshi:2008kw,Nakazato:2011vd,Peres:2012zt,Banik:2014rga,Chatterjee:2015pua,Raduta:2017wpp,Negreiros:2018cho,Grigorian:2018bvg,Raduta:2019rsk,Sedrakian2022,Malik:2022jqc,Sedrakian:2022ata,Fortin:2021umb}. We also note that recent studies have shown significant progress in measuring NS properties and determining the EoS partially employing statistical methods to combine different measurements and by this to decrease uncertainties (see e.g.~\cite{Antoniadis:2013pzd,Margalit:2017dij,Bauswein:2017vtn,Shibata:2017xdx,Ruiz:2017due,Radice:2017lry,Most:2018eaw,Rezzolla:2017aly,Koppel:2019pys,LIGOScientific:2018hze,Riley:2019yda,Miller:2019cac,Radice:2018ozg,Coughlin:2018fis,Dietrich:2020efo,Capano:2019eae,Riley:2021pdl,Miller:2021qha,Al-Mamun:2020vzu,Raaijmakers:2021uju,Breschi:2021tbm,Fonseca:2021wxt,Romani:2022jhd,Huth:2021bsp,Huang:2023grj}).

We already discussed above that the exact frequency shift will depend on the abundance of hyperons and thus on the threshold density for hyperon production. Smaller amounts corresponding to a high threshold density of hyperonization may not lead to strong effects and may thus remain undetected. The amount of hyperons produced will also depend on the mass of the system. One may generally expect that the impact of hyperons becomes more pronounced for more massive systems. In this work we have only considered a single system mass, since once $f_\mathrm{peak}$ can be measured with sufficient precision, the binary mass will be inferred with high accuracy. The increase of $f_\mathrm{peak}$ with the total binary mass can be estimated from Fig.~1 of Ref.~\cite{Bauswein:2014qla} showing that a mass uncertainty of about 0.1\% corresponds to a change of $f_\mathrm{peak}$ of a few~Hz, which is well below frequency shifts induced by the thermal behavior of hyperonic EoSs. We also emphasize that a frequency shift incompatible with purely nucleonic matter, does not necessarily indicate the presence of hyperons but more generally degrees of freedom that lead to a softening of the EoS. This includes in particular the possibility of deconfined quark matter as discussed in~\cite{Bauswein:2019skm,Weih:2019xvw,Bauswein:2020ggy,Blacker:2020nlq,Liebling:2020dhf,Prakash:2021wpz}. Pions may additionally affect the EoS~\cite{Vijayan:2023qrt}. Likely, additional information from either theory or experiments in the laboratory is essential to discriminate these possibilities. We also refer to recent studies indicating the possibility that the thermal behavior of purely nucleonic matter may yield reduced thermal pressure, i.e.~with a $\Gamma_\mathrm{th}$ significantly below 1.75, possibly even below 1 corresponding to negative thermal pressure~\cite{Keller:2020qhx,Keller:2022crb}. These state-of-the-art microscopic models cannot be used directly in simulations since they only produce results up to around two times nuclear saturation density (see~\cite{Bauswein:2010dn,Raithel:2021hye,Fields:2023bhs,Raithel:2023zml} for an exploration of thermal EoS effects in mergers). Phenomenological models developed to extrapolate these microscopic EoSs indicate that the thermal index reaches again larger values at higher densities~\cite{Huth:2020ozf}, which may lead to an average thermal index inside the star that is close to the usual nucleonic values adopted here. Furthermore, the inclusion of hyperons in nucleonic models with low average thermal index may still yield an additional softening of the thermal part of the EoS such that a very similar frequency shift relative to these nucleonic models may occur. We corroborate this argument with a toy model where we add hyperons to a model that we tuned such that the nucleonic part already features a drop in $\Gamma_\mathrm{th}$.

A solid interpretation of a possible frequency shift in a future detection will at any rate require a comprehensive comparison with advanced theoretical models and other insight from upcoming astronomical and laboratory measurements.

\section{Acknowledgments}
We thank K.~Chatziioannou for comments on the manuscript. H.K. and A.R. would like to thank Arnau Rios for useful discussions. We acknowledge funds by the State of Hesse within the Cluster Project ELEMENTS supporting H.K.'s visit during which this project was initialized. 
This research has been supported from the projects CEX2019-000918-M, CEX2020-001058-M (Unidades de Excelencia ``Mar\'{\i}a de Maeztu"), PID2019-110165GB-I00 and PID2020-118758GB-I00, financed by the spanish MCIN/ AEI/10.13039/501100011033/, as well as by the EU STRONG-2020 project, under the program H2020-INFRAIA-2018-1 grant agreement no. 824093, and by PHAROS COST Action CA16214. H.K. acknowledges support from the PRE2020-093558 Doctoral Grant of the spanish MCIN/ AEI/10.13039/501100011033/. 
S.B. and A.B. acknowledge support by Deutsche Forschungsgemeinschaft (DFG, German Research Foundation) through Project-ID 279384907 -- SFB 1245 (subproject B07). A.B. acknowledges support by the European Research Council (ERC) under the European Union’s Horizon 2020 research and innovation program under grant agreement No. 759253 and ERC Grant HEAVYMETAL No. 101071865 and support by the State of Hesse within the Cluster Project ELEMENTS.
L.T. also acknowledges support from the Generalitat Valenciana under contract PROMETEO/2020/023, from the Generalitat de Catalunya under contract 2021 SGR 171, and from the CRC-TR 211 'Strong-interaction matter under extreme conditions'- project Nr. 315477589 - TRR 211.

\input{main.bbl}
\end{document}

%% file: journalheader.tex
\newcommand{\aap}{Astron. Astrophys.}%
\newcommand{\mnras}{Mon. Not. Roy. Astron. Soc.}%